\documentclass[usenatbib]{mn2e} \usepackage{psfig}\usepackage{graphicx}\usepackage{amsmath}\usepackage{amssymb}

\title[Searching for hidden Wolf-Rayet stars in the Galactic Plane]
      {Searching for hidden Wolf-Rayet stars in the Galactic Plane --
      15 new Wolf-Rayet stars}

\author[Lucy.\,J.\,Hadfield et al.]
  {L.\,J.\,Hadfield$^{1}$\thanks{E-mail: l.hadfield@sheffield.ac.uk},
  S.\,D.\,Van\,Dyk$^{2}$, P.\,W.\,Morris$^{3}$, J.\,D.\,Smith$^{4}$,
  A.\,P.\,Marston$^{5}$ \newauthor and D.\,E. Peterson$^{6}$
  \\$^{1}$Department of Physics and Astronomy, University of
  Sheffield, Sheffield, S3 7RH, UK\\$^{2}${\it Spitzer Science
  Center}, IPAC, California Institute of Technology, M/C 220-6,
  Pasadena CA 91125, USA\\$^{3}$NASA Herschel Science Center, IPAC,
  California Institute of Technology, M/C 100-22, Pasadena CA 91125,
  USA\\$^{4}$Steward Observatory, University of Arizona, Tucson, AZ
  87512, USA\\$^{5}$ESA/ESAC, Villafranca del Castillo, P.O Box
  Apdo. 50727, 28080 Madrid, Spain.\\$^{6}$University of Virginia,
  Department of Astronomy P.O. Box 400325, Charlottesville, VA
  22904-4325, USA
}

\date{}

\begin{document}
  \maketitle
  
\begin{abstract}  
We report the discovery of fifteen previously unknown Wolf-Rayet
(WR) stars found as part of an infrared broad-band study of
candidate WR stars in the Galaxy.  We have derived an
empirically-based selection algorithm which has selected $\sim$5000 WR
candidate stars located within the Galactic Plane drawn from the GLIMPSE (mid-infrared) and 2MASS (near-infrared) catalogues.  Spectroscopic
follow-up of 184 of these reveals eleven WN and four WC-type WR stars.
Early WC subtypes are absent from our sample and none show evidence
for circumstellar dust emission.
Of the candidates which are not WR stars, $\sim$120 displayed hydrogen
emission line features in their spectra. Spectral features suggest
that the majority of these are in fact B supergiants/hypergiants,
$\sim$40 of these are identified Be/B[e] candidates.

Here, we present the optical spectra for six of the newly-detected WR
stars, and the near-infrared spectra for the remaining nine of our sample.
With a WR yield rate of $\sim$7\% and a massive star detection rate of
$\sim$65\%, initial results suggest that this method is one of the
most successful means for locating evolved, massive stars in the
Galaxy.

\end{abstract}



\section{Introduction}
\label{introduction}

Wolf-Rayet (WR) stars -- the chemically evolved descendants of the
most massive stars (i.e. $\geq$ 25\,M$_{\sun}$) -- possess intense
stellar winds and so continually enrich their local environments.
Although rare in number, high mass-loss rates during the WR stage
means that they can completely dominate the energetics and chemical
enrichment of their local interstellar medium (ISM). 
For example, WR stars comprise only 10\% of the massive stellar
content of NGC\,3603, but they are responsible for 20\% of the
ionizing photons and $\sim$60\% of the kinetic energy released into
the ISM \citep{pac98a}.  WR stars have also been suggested as possible
precursors of core-collapse supernovae of types Ib and Ic and
$\gamma$-ray bursts. 

Our Galaxy provides an excellent laboratory for studying massive
stars, as we can resolve objects on small scales and so hope to
achieve sample completeness. However, to date $\sim$300 WR stars have been
observed in our Galaxy \citep{derhucht06}, $\sim$75 of which have been
reported in the last five years.  With 
$\sim$1000--2500 WR stars expected to be located within the Milky Way
\citep{shara99, derhucht01}, it is highly likely that a large number of WR
stars are still waiting to be uncovered.

Previous WR surveys have utilised optical narrow-band interference
filters tuned to the strong, characteristic $\lambda$4686\,\AA\
He\,{\sc ii}/C\,{\sc iii} emission feature. These have proved to be
extremely successful, with the last large-scale optical survey
reporting 31 new WR stars \citep{shara99}.  However, the Galactic
Plane contains large amounts of gas and dust, such that high
extinction means that many WR stars may remain ``hidden'' from
view. Even if optical surveys were to go as deep as {\it V}=25\,mag,
it is estimated that one-third of the Galaxy's WR population would
still remain undetected \citep{shara99}.  As a result one must turn to
longer, infrared (IR) wavelengths in an attempt to reveal this
``missing'' population.
IR surveys have generally followed a narrow-band imaging approach in order to
detect WR stars.  WR features are much weaker at near-IR wavelengths,
but for locating WR stars in stellar clusters this method has proved
to be very successful \citep{pac06}.  Unfortunately, for large scale
surveys, especially those concentrated towards the Galactic Centre,
the observed success rate is very low.  \citet{homeier03} confirmed
only 4 new WR stars found using IR narrow-band
images.  With a lack of suitable filters for large-scale survey
instruments (e.g., WFCAM on UKIRT), such surveys require large
amounts of telescope time and produce large amounts of
data. 
With many known WR stars found to be isolated
in the field, it would appear that a significant number of WR stars
may still be waiting to be found.

\citet{williams81} and \citet{vandyk06} have showed that it is
possible to distinguish WR stars via their broad-band near-IR colours.
A combination of continuum and line emission properties of WR
stars distinguishes them from other stellar populations at IR
wavelengths, but as \cite{vandyk06} have found, they are not so
distinct that a criterion devised for candidate selection would lead
to an acceptable (in terms of telescope time) predicted success rate.
After confirming these suspicions in a pilot spectroscopic observing
run, they incorporated {\it MSX} 8.0\micron\ data to their colour selection
in order to broaden the colour baseline.  This resulted in the
discovery of one new WR star suggesting that a combination of near-
and mid-IR data can potentially provide a better discriminant for WR
stars.

Taking this scheme one step further, we have developed broad-band,
IR colour selection criteria which exploits 2MASS (near-IR) and the
mid-IR {\it Spitzer} GLIMPSE \citep[Galactic Legacy Infrared Mid-Plane
Survey Extraordinaire,][]{benjamin03} surveys\footnote{GLIMPSE
enhanced products are available at
http://data.spitzer.caltech.edu/popular/glimpse/}.  GLIMPSE imaged the
inner regions of the Milky Way within $\pm$1$^{\circ}$ of the Galactic
Plane at 3.6, 4.5, 5.8 and 8.0\,\micron\ with a sensitivity and
resolution that excellently matches that of 2MASS \citep[for a more
detailed description of the GLIMPSE project see][]{benjamin03}.
Covering only 220 square degrees of the Galactic Plane, the merging of
GLIMPSE and 2MASS colours excludes the possibility of an all plane
survey, but well complements previous WR searches, allowing progress
potentially to be made in the search for ``hidden'' WR stars in our
Galaxy.

This paper presents the discovery of 15 new WR stars confirmed via
either optical or near-IR spectroscopic follow-up observations.  It
complements a similar study conducted by \citet[][in
prep]{marston06}  who report the discovery of six new WR stars.
This paper is structured as follows: Sections~2 and 3 briefly summarise
the empirical colour selection algorithm used for target selection and
information regarding spectroscopic follow-up observations.  The newly
discovered WR stars are presented in Section~4, where we consider
spectral classification and provide reddening and distance estimates
for each star.  Although the primary goal of this survey was the
detection of WR stars, we have identified a large number of
emission-line objects.  We briefly discuss the non-WR detections in
Section~\ref{non:WR}.  Finally, we summarise our results and draw
conclusions in Section~\ref{conclusions}.

\section{Candidate selection}

In the following section we present a brief outline of the selection
criteria used to generate an initial candidate list. 

At IR wavelengths WR stars exhibit free-free emission characteristic
of a dense stellar wind \citep{wright75}.  This emission is
superimposed upon the stellar continuum such that WR stars exhibit a
flatter spectral energy distribution relative to normal early-type
stars.  In a future paper we will describe in detail the mechanisms
which drive the population of WR stars and subclasses to exhibit their
near- to mid-IR colours, following the phenomenology described by
\citet{vandyk06b}.

We have considered the positions of the observed colours of known
Galactic WR stars in all possible mid- and near IR colour-colour
combinations (i.e., GLIMPSE only, 2MASS only and GLIMPSE + 2MASS) to
investigate if it is possible to distinguish WR stars via their
broad-band IR colours.  We find that the known WR stars are best
distinguished in two possible colour-colour combinations.

\begin{figure*}
\begin{tabular}{l@{\hspace{-7mm}}l}
\psfig{figure=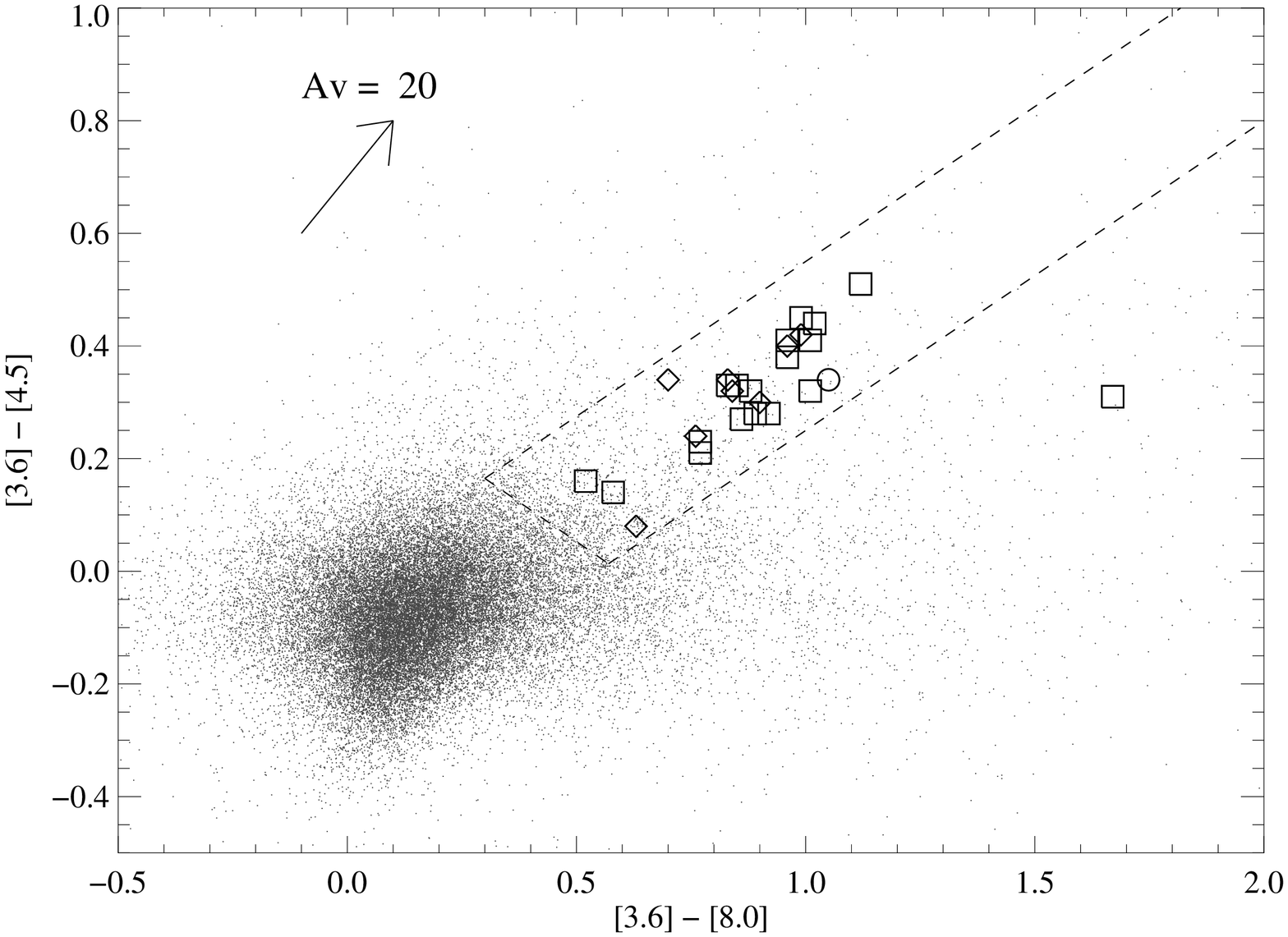,width=9.5cm,angle=90.}&\psfig{figure=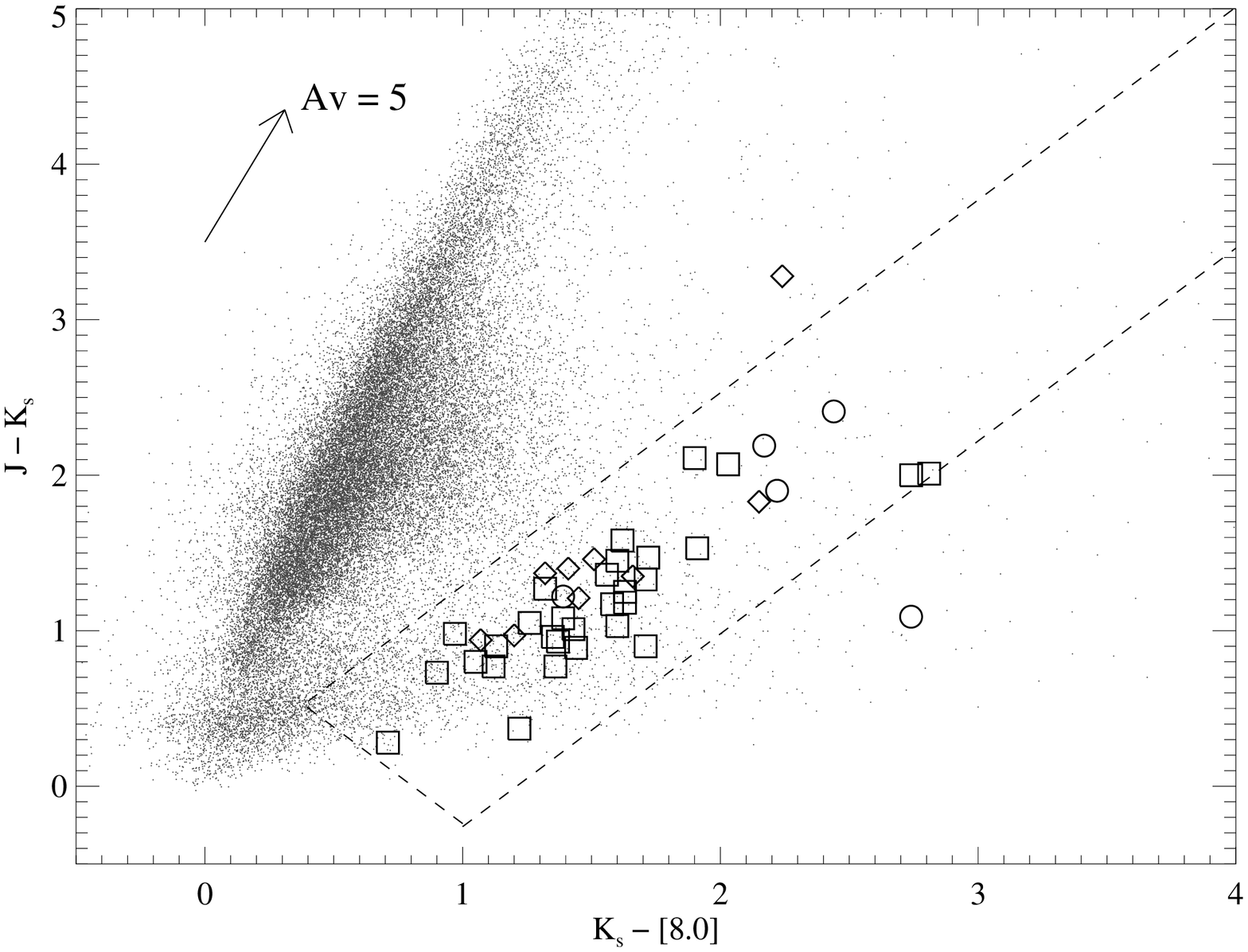,width=9.5cm,angle=90.}\\
\end{tabular}
\caption{IR Colour-Colour diagrams for a 1$^{\circ} \times$
  1$^{\circ}$ degree of the Galactic Plane centred on {\it
  l}=312$^{\circ}$ ({\it b}$\pm$1$^{\circ}$).  Previously reported
  Galactic WR stars (open symbols) clearly separate from the general
  stellar locus (points), and have been grouped into WN (squares), WC
  (diamonds) and WC9d (circles) stars.  The dashed lines indicate the
  limits used over all GLIMPSE fields to select targets for
  spectroscopic follow-up.  For a few cases, we find that their
  observed IR colours do not fall within the limits of our selection
  criteria.  However, all of these are located in complex environments
  which would suggest unreliable photometry for these stars. }
\label{fig:colourplots}
\end{figure*}

\begin{enumerate}
\item{{\bf Mid-IR colours:} Figure~\ref{fig:colourplots}(a) shows the
  mid-IR colours for stars contained in a 1$^{\circ} \times$
  1$^{\circ}$ degree of the Galactic Plane centred on {\it
  l}=312$^{\circ}$ ({\it b}$\pm$1$^{\circ}$).  Overplotted are the
  observed mid-IR colours of the known WR stars,
  which clearly separate from the general stellar locus. WR stars
  appear to exhibit a mid-IR excess, with colours [3.6] -- [8.0] $>$
  0.5 and [3.6] -- [4.5] $>$ 0.1.  WN and WC stars appear to exhibit
  similar colours, such that mid-IR colours alone do not differentiate
  between WR subtypes.}
\item{{\bf Mid + near-IR colours:} Galactic WR stars also appear to be
  distinguished in a combination of near and mid-IR colours.  The 2005
  ``reliable'' GLIMPSE catalogue was cross-matched with the 2MASS
  all-sky point source catalogue using a search radius of 1\arcsec\ to
  create the colour sequence shown in Figure~\ref{fig:colourplots}(b). 
  As expected, the majority of the objects follow the reddening
  vector, while the WR stars clearly exhibit an 8.0\micron\ excess.
  The WR stars appear to follow a separate, well-defined path relative
  to the reddening vector. Again WN and WC stars exhibit similar
  colours, however, WC stars with unresolved circumstellar dust shells
  have a tendency to display the largest 8.0\,\micron\ excess.  }
\end{enumerate}

%
Using the positions of the known WR stars we have outlined two regions
of colour space (as indicated in Fig~\ref{fig:colourplots}) which
define our WR candidates.  Of the
$\sim$2$\times 10^{7}$ objects common to both to the GLIMPSE and
2MASS catalogues, $\sim$100\,000 were initially selected as candidate WR stars.

To reduce this number of candidates we have imposed additional
selection criteria and restricted ourselves to only the most reliable
sources, i.e., objects that are unblended, unconfused and of the
highest photometric quality. This does, of course introduce a sample
bias against highly crowded regions, i.e., stellar clusters.  In
addition to these ``quality flags'', we have applied a 2MASS only
(J--H vs H--K$_{S}$) colour selection criterion using the WR near-IR
colour relation reported by \citet{vandyk06}. Additionally, assuming
that optical WR surveys are complete to {\it B}=14\,mag
\citep{shara99}, the final selection criterion was to place a B$>$14
magnitude limit for all sources which have optical counterparts within
a 3\arcsec\ radius listed in the USNO-A2.0 catalogue.

Including all of the above restrictions the final number of WR
candidates was reduced to $\sim$5700. Candidates were checked against
known objects listed in the CDS SIMBAD database
(http://simbad.u-strasbg.fr/), within a 10\arcsec\ radius. Objects which
already had spectroscopic classifications were excluded from our
sample ($\sim$1\%).  An additional $\sim$5\% were listed as possible
emission line objects or variable stars.  The rest had no
identification in SIMBAD.
 
In order to observe a representative sample for each longitude bin,
targets were grouped by magnitude, and a subsample was selected at
random from each magnitude bin.  Objects that appeared to be extended
in 2MASS or Digitized Sky Survey (DSS) images were excluded.

\begin{table*}
\caption{Observing Log}
\begin{tabular}{lllccccc}
\hline
ID&Date&\multicolumn{1}{c}{Observations}&Slit&R&Wavelength&Candidates&WR\\
&&&(\arcsec)&&\multicolumn{1}{c}{Coverage}&\multicolumn{1}{c}{Observed}&\multicolumn{1}{c}{Discoveries}\\
\hline (1)&2006 March
19--24&4.0m-Blanco/RCspec&1.0&900&3900--7500\AA&97&6\\ (2)&2006 June
10--11&3.5m-ARC/CorMASS&2.0&300&0.8--2.4\micron&26&2\\ (3)&2006 August
8--9, 11 &3.0m-IRTF/SpeX&0.8&750&0.8--2.4\micron&41&6\\ (4)&2006
September 4--5&3.0m-IRTF/SpeX&0.3&2000&0.8--2.4\micron&20&1\\\hline
\end{tabular}
\label{observations}
\end{table*}    

\section{Spectroscopic Observations}
\label{sec:Observations}
The discovery of the fifteen new WR stars presented here is the
result of four spectroscopic observing campaigns. Optically bright WR
candidates in the Galactic latitude range of {\it
l}=284$^{\circ}$--350$^{\circ}$ were observed from the Cerro Tololo
Inter-American Observatory (CTIO) on the nights of 2006 March
19--24. Near-IR spectroscopy of candidates within the GLIMPSE range
{\it l}=0--65$^{\circ}$ was obtained from the Apache Point
Observatory (APO) 2006 June 10--11 and the Infrared Telescope Facility
(IRTF), Mauna Kea on 2006 August 8--9 and 11 and September 4--5.  For
classification purposes we have also observed a sample of known WR
stars and transition stars.  An observational log is presented in
Table~\ref{observations}.

\subsection{Optical Spectroscopy}   

Long-slit spectroscopic observations of 97 of our southern hemisphere
candidates were obtained at the CTIO 4m-Blanco Telescope with the
RCSpec spectrograph. Seeing conditions were $\sim$0.5\arcsec--1.0\arcsec, and
all targets were observed at an airmass $\lesssim$1.6.  Exposure
times ranged between 300\,s for our brighter targets ({\it
B}$\sim$15\,mag), to 1800\,s for the fainter objects ({\it
B}$\sim$19\,mag).  The data were obtained using the KPGL3 grating with
a 1\arcsec\ slit, resulting in a dispersion of $\sim$1.2\,\AA/pixel and
resolution of 3.8\,\AA, as measured from comparison arc lines.  The
wavelength range of the final spectra was $\sim$4000--7500\AA.
 
The data were cleaned of cosmic ray hits, de-biased and divided by a normalised
flat field.  Spectra were extracted using standard reduction routines
within {\sc iraf}.  Bad rows and pixels were masked assuming that they
corresponded to $\sim$10$\sigma$ outliers on flat-field images; any
additional flaws were identified using the spectrum of the photometric
standard star LTT\,62648.  Wavelength calibration arcs were taken at
regular positions across the sky  to allow for instrument flexure.
Atmospheric absorption features have not been removed from the final
spectra.

The majority of targets showed evidence of high extinction as the
signal-to-noise ratio ({\it S/N}) of the spectra rapidly deteriorated
at shorter wavelengths.  The blue ($\sim$4500\AA) continuum {\it S/N}
ranged from $\sim$1--30, versus $\sim$30--100 in the red
($\sim$6800\AA).  Nevertheless, blue WR features were typically
detected at the 25$\sigma$ level.

\subsection{Near-IR Spectroscopy}
\label{IR:Observations}
Near-IR spectroscopic follow-up observations of 87 candidate WR stars
were acquired using the ARC 3.5-m with CorMASS \citep[Cornell
Massachusetts Slit Spectrograph,][]{wilson01} and the IRTF 3.0-m with
SpeX \citep{rayner03}.  

CorMASS and SpeX are both near-IR cross-dispersed spectrographs.
CorMASS is a low resolution detector and has a fixed slit width of
$2{\farcs}0$.  SpeX is a medium resolution and was used in the
short-wavelength cross-dispersed (SXD) mode with a $0{\farcs}8$ and
$0{\farcs}3$ slit, resulting in R=750 and 2000, respectively.  All
near-IR observing modes resulted in a spectral coverage of 0.8--2.4\micron.

In order to maximise the number of candidates observed, a limiting
magnitude of K$\sim$10.5 and K$\sim$11.0\,mag was imposed for Spex and
CorMASS spectroscopy, respectively. 
  
The Spex and CorMASS data were reduced using the Interactive Data
Language reduction software tools Spextool\,v3.3 and Cormasstool\,v3.4
\citep{cushing04}.  As the data were acquired using the standard
``nod'' technique, individual 2D frames were divided through by a
normalised flat field before each pair was subtracted to produce a
background-subtracted image; any residual background was subsequently
removed during the extraction process.  Spectra were extracted using
an appropriate aperture radius, which for the majority of sources was
$\sim$1\arcsec.

Atmospheric features were removed from the final spectra by observing
a selection of A0V standard stars following \citet{vacca03}. The
near-IR spectra of the nine of the bona-fide WR stars are shown in
Figure~\ref{fig:spectra_ir}.


\section{Results: 15 new WR stars}

Of the 184 candidates spectroscopically observed we have confirmed the
discovery of 15 previously unknown WR stars.  For brevity we have
assigned a HDM nomenclature to these new discoveries.  2MASS
designations and magnitudes of these new discoveries are presented in
Table~\ref{tab:mag}.  K$_{S}$-band finding charts can be found in
Appendix~\ref{findingcharts}.

\begin{table*}
\caption{Data for the newly discovered Wolf-Rayet stars found by our
  survey. 
  {\it JHK$_{s}$} magnitudes are taken from the 2MASS
  catalogue and mid-IR magnitudes correspond to values listed in the 2005
  reliable GLIMPSE catalogue.  Blue and red magnitudes are those
  listed in the 2MASS catalogue and are taken from the USNO-A2.0
  Catalogue.}
\begin{tabular}{lcrrcrrrlllllc}
\hline
Star& 2MASS &\multicolumn{1}{c}{{\it l}} & \multicolumn{1}{c}{{\it b}} &B & R&\multicolumn{1}{c}{J} & \multicolumn{1}{c}{H} & \multicolumn{1}{c}{K$_{s}$} & [3.6] & [4.5] &[5.4] & [8.0]& Obs\\
  &Designation & &&& & & & &&&&&ID\\
\hline
HDM\,1&11423766--6241193 &295.13 & --0.86&18.5 & 16.6 &11.50 & 10.54 &  9.81 &  9.07 & 8.66 & 8.42 & 7.60&(1) \\
HDM\,2&  12065647--6238304 & 297.85&--0.21&17.8 & 15.3 & 11.55 & 10.82 & 10.22 &  9.56 & 9.19 & 8.99 & 8.59&(1)  \\
HDM\,3& 12133878--6308580 &298.68&--0.59&17.7 & 14.5& 10.56 &  9.64 &  9.03 &  8.25 & 7.99 & 7.74 & 7.35  &(1) \\
HDM\,4&12461614--6257234 &302.34&--0.09& 19.6 & 16.5 & 12.18 & 11.22 & 10.56 &  9.81 & 9.43 & 9.21 & 8.27&(1)  \\
HDM\,5& 13101207--6239065 & 305.08 &+0.14& 18.1 & 15.7&  11.06 &  10.09 &  9.32 &  8.59 & 8.32 & 8.13 & 7.79& (1) \\
HDM\,6&16113927--5205458&331.03&--0.50& 18.3 & 15.3 &10.16&9.25&8.46&7.86&7.48&7.36 & 6.89 &(1) \\
HDM\,7&18094505--2017103&10.23&--0.41&-- &--&12.58&11.06& 10.14& 9.10&  8.74&  8.49&  8.07&(3)\\
HDM\,8&18131420--1753434&12.72&+0.02&19.2&15.2&9.62&8.60&9.94&7.32&6.87&6.71&6.34&(3)\\
HDM\,9&18192219--1603123&15.04&--0.39&16.5&14.3&9.09&8.30&7.76&7.13&6.80&6.65&6.22&(3)\\
HDM\,10&18250024--1033236&20.53&+0.98&--&--&12.21& 11.24& 10.61&9.74&  9.33&  9.03&  8.67&(4)\\
HDM\,11&18255310--1328324&18.05&--0.57&-- &--& 10.32 & 9.52 & 8.96 & 8.27 & 7.95 & 7.79 &7.35&(3)\\
HDM\,12&18400865--0329311&28.54&+0.92&-- &--&11.77& 10.86& 10.30&9.54&9.28&  8.99&  8.64&(2)\\
HDM\,13&18411070--0451270&27.44&+0.06&-- &--&11.05& 10.17&  9.45& 8.80&  8.32& 8.20& 7.74&(3)\\
HDM\,14&19081797+0829105&42.40&+0.15&--&--&10.59&9.47&8.71&8.09&7.60&7.38&6.97&(2) \\
HDM\,15&19334401+1922475&54.91&--0.16&16.3&13.7&10.20&9.61&9.07&8.13&7.75&7.48&7.01&(3)\\
\hline
\end{tabular}
\label{tab:mag}
\end{table*}


\subsection{Spectral Classification}

Classification of WR stars via their optical spectra is relatively
straightforward as schemes using optical line diagnostics are well
defined \citep{smith96, pac96}.  In the near-IR, classification is
somewhat more difficult as there are no line diagnostics which
uniquely distinguish all WR subtypes \citep{figer97, pac06}.  Optical
and near-IR classification will therefore be discussed separately.

\subsubsection{Optical Classification}

The reduced optical spectra (normalised to the continuum) are
presented in Figure~\ref{fig:spectra}, grouped into WN and WC
subtypes. Prominent WR emission features are identified.  In order
to classify the newly discovered WR stars, we have fit Gaussian line
profiles to the prominent WR features.  Measured
equivalent widths and FWHM of the prominent optical WR features are presented
in Table~\ref{tab:prop}.

\begin{table*}
\caption{Emission line equivalent widths (W$_{\lambda}$) and FWHM
  (both in \AA) of prominent optical WR emission features.  WR
  subtypes have been assigned using the classification schemes of
  \citet{smith96} and \citet{pac98}.}
\begin{tabular}{lc@{\hspace{1mm}}cc@{\hspace{1mm}}cc@{\hspace{1mm}}cc@{\hspace{1mm}}cc@{\hspace{1mm}}cc@{\hspace{1mm}}cc@{\hspace{1mm}}cl}
\hline Star & \multicolumn{2}{c}{N\,{\sc iii}/C{\sc iii}} & \multicolumn{2}{c}{He\,{\sc ii}}&
\multicolumn{2}{c}{He\,{\sc ii}}
&\multicolumn{2}{c}{C{\sc iii}}
&\multicolumn{2}{c}{C{\sc iv}}
&\multicolumn{2}{c}{He\,{\sc ii}}
&\multicolumn{2}{c}{H\,$\alpha$}&Spectral\\
&\multicolumn{2}{c}{(4640/50)}&\multicolumn{2}{c}{(4686)}&\multicolumn{2}{c}{(5411)}&\multicolumn{2}{c}{(5696)}&\multicolumn{2}{c}{(5808)}&\multicolumn{2}{c}{(5876)}&\multicolumn{2}{c}{(6560)}&Type\\
&FWHM&W$_{\lambda}$&FWHM&W$_{\lambda}$&FWHM&W$_{\lambda}$&FWHM&W$_{\lambda}$&FWHM&W$_{\lambda}$&FWHM&W$_{\lambda}$&FWHM&W$_{\lambda}$&\\
\hline 
HDM\,1&9&10&9&10&--&--&7:&2:&--&--&7&30&9&85&WN9--10h\\
HDM\,2&30&47&22&132&25&25&--&--&26&25&22&13&28&56&WN6o\\ 
HDM\,3&20&17&14&33&11&5&--&--&weak&weak&15&7&24&57&WN7h\\ 
HDM\,4&29&41&22&118&21&30&--&--&27&28&21&10&26&65&WN6o\\ 
HDM\,5&25&343&25&119&25&30&37&83&42&212&31&289&&&WC7\\ 
HDM\,6&19&107&21&42&12&10&21&68&24&292&22&67&&&WC9\\ 
\hline
\end{tabular}
\label{tab:prop}
\end{table*}

\begin{figure}
\centerline{\psfig{figure=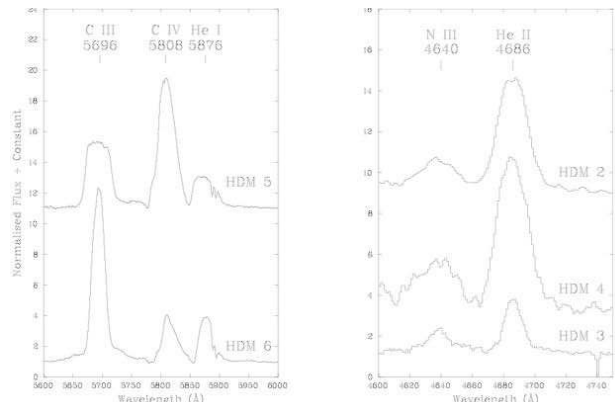,bbllx=40pt,bblly=30pt,bburx=550pt,bbury=850pt,clip=,width=\columnwidth,angle=-90.}}
\caption{Optical CTIO/Rcspec spectra of the prominent yellow WC
  features (left) and blue WN (right) features. }
\label{fig:optical}
\end{figure}

As Fig~\ref{fig:optical} shows, HDM\,5 and 6 both display strong
carbon features associated with late-type WC subtypes.  Following the
quantitative WC classification scheme of \citet{pac98}, refined from
that of \citet{smith90}, we assign WC7 and WC9 subtypes to HDM\,5 and
HDM\,6, respectively.

Late-type WC stars have the potential of being surrounded by
circumstellar dust shells which contribute thermal emission to
the stellar continuum at mid-IR wavelengths, depending on dust
temperature and density.  Using optical spectroscopy it is not
possible to directly confirm the presence of dust.  However, both
HDM\,5 and 6 display near-IR colours consistent with non-dusty WC
stars.

The remaining four WR spectra display He and N emission features
indicative of WN stars.  For this analysis we have adopted the
classification scheme of \citet{smith96} and the observed He\,{\sc i}
$\lambda$5876 / He\,{\sc ii} $\lambda$5411 line ratios indicate a WN7
subtype for HDM\,3, and WN6 for both HDM\,2 and 4.

\begin{figure}
\centerline{\psfig{figure=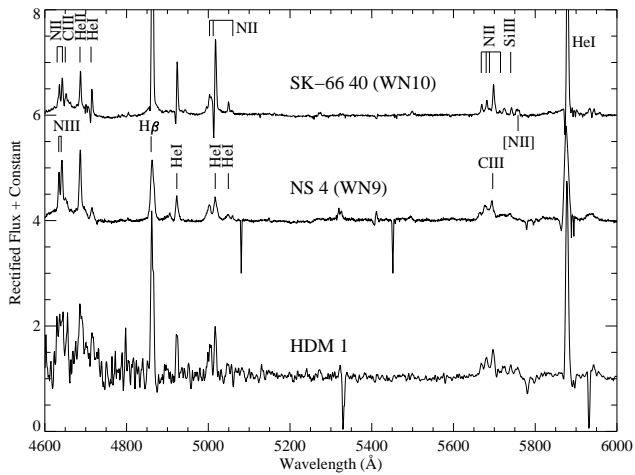,bbllx=60pt,bblly=60pt,bburx=540pt,bbury=700pt,clip=,width=\columnwidth,angle=90.}}
\caption{Comparison of HDM\,1, NS4/WR105 (WN9) and LMC star Sk--66\,40 (WN10).
  The spectra of NS4/WR105 and SK-66\,40 are taken from \citet{bohannan99}
  and \citet{pac95a}.  }
\label{fig:WNL}
\end{figure}

The spectrum of HDM\,1 exhibits distinct differences to
the other WN stars in our sample.  The spectral features are notably
narrower (FWHM$\sim$7\AA) and lower excitation features are present.
We compare the blue ($\lambda\lambda$4600--5000) and yellow
($\lambda\lambda$4600--5000\AA) spectrum (normalised to the continuum)
of HDM\,1 with that of Galactic star NS4/WR105 (WN9) and LMC star
Sk--66\,40 (WN10) in Figure~\ref{fig:WNL}.  The ``blue WR bump''
(He\,{\sc ii}, N\,{\sc ii--iii} and C\,{\sc iii}) is clearly present, as
are yellow N\,{\sc iii}, C\,{\sc iii} and Si\,{\sc iii} features.  The
spectral appearance of HDM\,1 is remarkably similar to that of the
comparison stars, and from that standpoint we classify HDM\,1 as
WN9/10.  Spectral types  WN9--10 are distinguished using the
relative strengths of the blue N\,{\sc ii}--{\sc iv} emission lines
\citep{smith96,pac97}.  Poor blue {\it S/N} prevents a classification
based on this method for our object, so we have used the He\,{\sc
ii}\,$\lambda$4686 vs. the He{\sc i}\,$\lambda$5876 lines as a subtype
indicator \citep{pac97}. This favours a WN9 subtype for HDM\,1, but an
accurate classification requires further observations.

WN stars can be further classified as weak or broad lined stars based
on the FWHM of the He\,{\sc ii} spectral features.  A star is
considered to be broad lined if FWHM (He\,{\sc ii}\,4686) $\geq$30\AA\
and/or FWHM (He\,{\sc ii}\,5411) $\geq$40\AA\ \citep{smith96}.
All of the WN stars for which optical spectroscopy is available are
consistent with weak/narrow lined WN stars.

WR stars are normally hydrogen deficient objects, so the presence of
hydrogen in WN stars has an important role from an evolutionary
standpoint.  A detailed abundance determination would, of course,
require a quantitative spectral analysis, but it is possible to assess
the presence of hydrogen using Pickering and Balmer lines at optical
wavelengths.

Using blue He\,{\sc ii} / He\,{\sc ii} + H$\beta$ line ratios,
\citet{smith96} denote the presence of hydrogen with an additional 
o, (h), h, classification to signify no, weak or strong hydrogen.
For our sample, poor S/N data prevents a reliable analysis of He\,{\sc
ii} $\lambda 4541$ and so we have qualitatively assessed our spectra
using He\,{\sc ii} + H$\beta$ ($\lambda 4860$), He\,{\sc ii}+
H$\alpha$ ($\lambda 6560$) and He\,{\sc ii} ($\lambda$5411).  Only for
HDM\,1 and 3 are the hydrogenic He\,{\sc ii} lines strong relative to
He\,{\sc ii} $\lambda 5411$, suggesting a WN9--10h and WN7h
classification for these objects.


\subsubsection{Near IR Classification}

Recently, with reference to optical line diagnostics (i.e., helium and
carbon lines), \citet{pac06} presented a near-IR classification scheme
which showed that these line diagnostics serve as reliable subtype
discriminators.  For stars earlier than WN6 or WC7, classification
requires additional nitrogen and carbon diagnostics.  Here, we have a
adopted the classification scheme of \citet{pac06} as well as making
spectral comparisons with previously known WR stars.  Line strengths
of the prominent near-IR WR features are summarised in
Table~\ref{tab:propir}.

\begin{table*}
\caption{Emission line equivalent widths (W$_{\lambda}$, in \AA) of
  near-IR emission line diagnostics used to classify the new WR
  stars.  WR subtypes have been assigned using the classification
  scheme of \citet{pac06}.$^{*}$ indicates that it was not possible to
  de-blend the spectral feature and values quoted refer to the strength
  of the whole feature.}
\begin{center}
\begin{tabular}{lccccccl}
\hline
Star&He\,{\sc ii}&He\,{\sc i}&N\,{\sc v}&N\,{\sc iii}/He\,{\sc ii}&He\,{\sc ii}+Br$\gamma$&He{\sc ii}&Spectral\\
&(1.012$\mu$m)&(1.083$\mu$m)&(2.110$\mu$m)&(2.115$\mu$m)&(2.165$\mu$m)&(2.189$\mu$m)&Type\\
\hline
HDM\,7&221&187&40&90&7&31&WN6o\\
HDM\,8&132&338&10&45&59&62&WN7b\\
HDM\,9&54&244&52&32&3&38&WN7o\\
HDM\,10&230$^{*}$&720$^{*}$&9&65&87&77&WN7b\\
HDM\,11&194&207&42&84&6&30&WN6o\\
HDM\,12&69&350&58&28&5&49&WN7o\\
HDM\,14&262&262&55&102&9&34&WN6o\\
\hline
&C\,{\sc iii}&C\,{\sc ii}&C\,{\sc iv}&C\,{\sc iii}&C\,{\sc iv}&C\,{\sc iii}&\\
&(0.971$\mu$m)&(0.990$\mu$m)&(1.191$\mu$m)&(1.198$\mu$m)&(2.076$\mu$m)&(2.110$\mu$m)&\\
\hline
HDM\,13&95&70&11&73&9&78&WC9\\
HDM\,15&363&19&51&42&270$^{*}$&95&WC8\\
\hline
\end{tabular}
\end{center}
\label{tab:propir}
\end{table*}

\begin{figure}
\psfig{figure=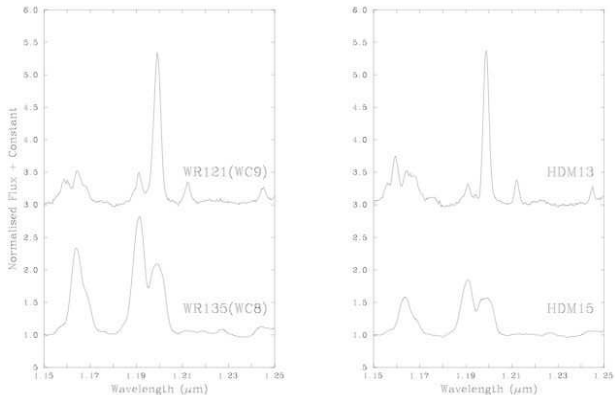,bbllx=30pt,bblly=20pt,bburx=550pt,bbury=850pt,clip=,width=\columnwidth,angle=-90.}
\caption{Spectral comparison of the near-IR C\,{\sc iii} / C\,{\sc iv}
  morphology of WR\,121 and WR\,135 with HDM\,13 and 15.  All spectra
  were acquired with the IRTF and SpeX.}
\label{fig:WCL}
\end{figure}

HDM\,13 and 15 both display strong near-IR carbon features and from
{\it z} and K-band C\,{\sc iv}/C\,{\sc iii} line ratios we infer WC9 and WC8
subtypes for HDM\,13 and 15.  These classifications are confirmed in
Figure~\ref{fig:WCL}, where we compare the C\,{\sc iv} 1.19 and C\,{\sc
iii} 1.20\,$\mu$m features of the newly discovered WR spectra with
those of well-classified WC stars.

The presence of circumstellar dust is easily inferred from near-IR
spectra as the WR K-band spectral features are diluted by dust
emission.  Of the newly presented WC spectra, none show evidence of
dust emission.

For HDM\,7--12 and HDM\,14 spectral features are consistent with
nitrogen-rich WR stars. He\,{\sc ii} 1.012$\mu$m / He\,{\sc i} 1.083$\mu$m and
He\,{\sc ii} 2.189$\mu$m / Br$\gamma$ line ratios suggest a WN7 subtype for
HDM\,8--10 and 12, whereas we infer a WN4--6 subtype for HDM\,7, 11 and 14.

\begin{figure}
\centerline{\psfig{figure=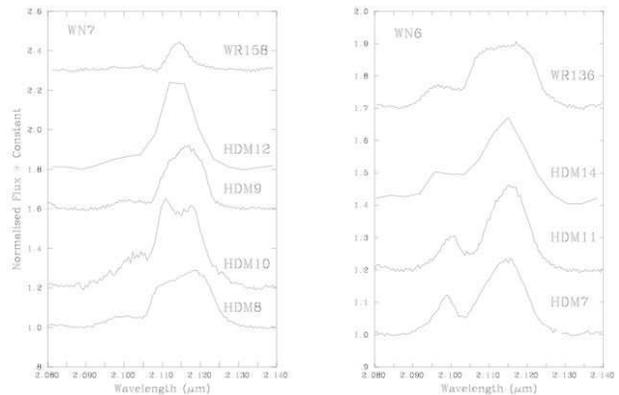,bbllx=30pt,bblly=20pt,bburx=550pt,bbury=850pt,clip=,width=\columnwidth,angle=-90.}}
\caption{A spectral comparison of the near-IR N\,{\sc iii} / N\,{\sc
  v} morphology of the newly discovered WN7 (left) and WN6 (right)
  stars with that of WR\,158 (WN7) and WR\,136 (WN6b).}
\label{fig:WN6}
\end{figure}

For WN4--6 stars, He line diagnostics do not provide a unique
discriminator and the spectral type must be further refined using the
2.11\,$\mu$m N\,{\sc iii} / N\,{\sc v} morphology.  For the stars
presented here N\,{\sc iii} is strong relative to N\,{\sc v}, and all
have N line ratios consistent with a WN6 subtype (see
Fig~\ref{fig:WN6}).

In the near-IR, WN stars are classified as broad-lined stars if the
FWHM\,(He\,{\sc ii} 2.1885$\mu$m) $>$ 130\,\AA\ \citep{pac06}.  Using
this criterion, HDM\,8 and HDM\,10 both qualify as being a broad-lined
WN, such that we assign WN7b to these stars.

The presence of hydrogen can be assessed from the near-IR spectra
using the He\,{\sc ii} 2.1889\,$\mu$m / Br$\gamma$ line ratios.
Observed line ratios for the WN6 stars mimic those of the H-deficient
WN6 stars in Westerlund~1 \citep{pac06}, such that we strongly suspect
that our WN6 stars are H-deficient.  The same conclusions can be drawn
for the narrow lined WN7 stars.


\subsection{Extinction and Distance}
\label{extinction}
WR photometry can be used to estimate the extinction and distance
towards our stars, providing we have a calibration of intrinsic
colours and absolute magnitudes. In principle, both visual and IR
wavelengths can be used to derive reddenings for our sample, but with
extinction at {\it JHK} wavelengths expected to be relatively
insensitive to the line-of-sight we have chosen to assess the
reddening and distances of our stars using their 2MASS {\it JHK$_{S}$}
colours \citep{Indebetouw05}.

Interstellar K-band absorption has been estimated by comparing the
observed and subtype-dependent intrinsic J--K$_{S}$ and H--K$_{S}$
colours.  Heliocentric distances have been derived using the mean
value of A$_{K_{S}}$.  The A$_{[\lambda]}$/A$_{K_{S}}$ ratios are
taken from \citet{Indebetouw05} and intrinsic WR photometric
properties are those listed in \citet{pac06}.  Intrinsic WR
colours correspond to ESO/SOFI {\it JHK$_{S}$} filters whereas
absolute magnitudes are based on 2MASS photometry. Since
SOFI and 2MASS have comparable filter passbands, there should be
negligible differences between SOFI and 2MASS intrinsic WR colours.
Galactrocentric radii (R$_G$) have been calculated assuming
R$_{\odot}$ = 8.5\,kpc. 

\begin{table}
\caption{Estimates of the extinction ($\overline{\mbox{A}_{K}}$) and
  distance (D) derived using 2MASS near-IR colours. Intrinsic colours and absolute K$_{S}$
  magnitudes are taken from \citet{pac06}. Galactrocentric radii (R$_G$) have been calculated
  assuming R$_{\odot}$ = 8.5\,kpc.}
\begin{tabular}{@{\hspace{1mm}}r@{\hspace{3.0mm}}l@{\hspace{1.5mm}}r@{\hspace{1.5mm}}r@{\hspace{1.5mm}}r@{\hspace{3.0mm}}r@{\hspace{3.0mm}}r@{\hspace{1mm}}r@{\hspace{0mm}}r@{\hspace{1mm}}}
\hline 
HDM& SpType&\multicolumn{1}{c}{K$_{S}$} & \multicolumn{1}{c}{A$_{K_{s}}^{J-K}$}
&\multicolumn{1}{c}{A$_{K_{s}}^{H-K}$}&$\overline{\mbox{A}_{K}}$& M$_{K_{s}}$&\multicolumn{1}{c}{D} & \multicolumn{1}{c}{R$_{G}$}\\
\hline 
1  &	WN9-10h & 9.81  &1.12 &  1.04 &  1.08 &-5.92&  8.5  & 9.1  \\
2  &	WN6o	& 10.22  &0.79 &  0.76 &  0.78 &-4.41&  5.9  & 7.8  \\
3  &	WN7h	& 9.03   &0.91 &  0.93 &  0.92 &-5.92&  6.4  & 7.8  \\
4  &	WN6o	& 10.56  &0.91 &  0.97 &  0.94 &-4.41&  6.4  & 7.4  \\
5  &	WC7	& 9.32   &0.36 &  0.75 &  0.55 &-4.59&  4.7  & 7.0  \\
6  &	WC9	& 8.46   &0.96 &  0.98 &  0.97 &-6.30&  5.7  & 4.5  \\
7  &	WN6o	& 10.14  &1.38 &  1.51 &  1.44 &-4.41&  4.2  & 4.5  \\
8  &	WN7b	& 7.94   &0.71 &  0.87 &  0.79 &-4.77&  2.4  & 6.2  \\
9  &	WN7o	& 7.76   &0.78 &  0.80 &  0.79 &-5.92&  3.8  & 5.0  \\
10 &	WN7b	& 10.61  &0.65 &  0.82 &  0.74 &-4.77&  8.5  & 3.0  \\
11 &	WN6o	& 8.96   &0.73 &  0.79 &  0.76 &-4.41&  3.3  & 5.4  \\
12 &	WN7o	& 10.30  &0.82 &  0.89 &  0.86 &-5.92&  11.8 & 6.0  \\
13 &	WC9	& 9.45   &0.84 &  0.91 &  0.87 &-6.30&  9.4  & 4.4  \\
14 &	WN6o	& 8.71   &1.09 &  1.13 &  1.11 &-4.41&  2.5  & 6.9  \\
15 &	WC8	& 9.07   &0.29 &  0.47 &  0.38 &-4.65&  4.7  & 7.0  \\
\hline	 
\end{tabular}
\label{tab:dist}
\end{table}

Derived K-band extinction and distance estimates for our WR sample are
summarised in Table~\ref{tab:dist}.  In general we find that
$\overline{\mbox{A}_{K_{S}}} \sim$0.8--1\,mag, except for HDM\,7 where
$\overline{\mbox{A}_{K_{S}}}$ is significantly larger.  This higher
extinction is supported by DSS  images which
shows no evidence of an optical counterpart.  All of the other WR
stars in our sample appear to have optical counterparts, even though
for some optical magnitudes are not listed in the USNO-A2.0 catalogue.

For the majority of our stars, A$_{K}^{J-K}$ and A$_{K}^{H-K}$ are in
good agreement (to within $\sim$0.1\,mag). 
WR stars are known to exhibit a wide range of intrinsic colours, and
it is this scatter that leads to large uncertainties in our derived
extinctions.  The colour--subtype scatter of the stars which
\citet{pac06} used to estimate intrinsic WR colours lead to 10\%
uncertainties in derived extinctions.

For HDM\,5, A$_{K}^{J-K}$ and A$_{K}^{H-K}$ differ by 0.4\,mag,
suggesting that our adopted WC7 intrinsic colours are unsuitable.  If
this star has intrinsic IR colours more consistent with those observed
for WR68 (i.e., $\left[J - K\right]_{o}$=0.24, $\left[H -
K\right]_{o}$=0.32, van der Hucht 2006), A$_{K}^{J-K}$ and
A$_{K}^{H-K}$ would be consistent to within 0.2\,mag, and
$\overline{\mbox{A}_{K}}$ would increase from 0.55 to 0.90\,mag.  This increase
in extinction  would reduce our estimated distance from 4.7\,kpc
to 4.0\,kpc.

Intrinsic absolute K$_{S}$-band magnitudes are also seen to display a
large scatter, even when considering individual WR subtypes.  With the
observed spread in M$_{\mbox{K}_{S}}$ being of the order of $\pm$0.5\,mag
\citep{pac06}, and assuming that this dominates the uncertainty in our
derived distances, we estimate that our heliocentric distances are
correct to within $\sim$25\%.

\section{Galactic WR distribution}
In order to investigate how the distribution of newly discovered WR
stars compares with that of the previously known Galactic WR stars it
is necessary to recompute distances to the WR stars listed in the
VIIth Catalogue of WR stars, and its Annex \citep{derhucht01,
derhucht06} using the intrinsic near-IR colour/magnitude calibrations
described in Section~\ref{extinction}.  For those WR stars which
belong to stellar clusters or OB associations, we have adopted
previously derived distances from the literature (see
\citet{derhucht01, pac06} and references therein). The only exception
is for the highly obscured cluster SGR 1806--20, where our
derived distance (based on a WN6 calibration) suggests a distance of
$\sim$6\,kpc rather than that of 15\,kpc proposed by
\citet{eikenberry04}.

For binary members, extinction estimates have been made assuming that
the WR star dominates the near-IR colours of the system. The absolute
magnitude of the system has been calculated using the synthetic O-star
photometry of \cite{martins06}.  Binarity was only considered for cases
where the companion star is well classified. 

Estimating the extinction towards dusty WC stars using near-IR
photometry is not straightforward since the associated circumstellar
dust can exhibit a wide range of properties. At optical wavelengths
the contribution of dust is negligible so that optically derived
extinctions should, in principle, be more realistic than those
estimated from IR colours. Therefore, solely for dusty WC stars, A$_K$ has
been estimated from optically derived extinctions listed in the VIIth
Catalogue assuming that A$_{V}$=8.8A$_{K}$
\citep{Indebetouw05}. Distances were then derived adopting
M$_{K_{S}}$=--8.5$\pm$1.5\,mag \citep{pac06}.  Applying this calibration to
WR95 yields a heliocentric distance of 3.9\,kpc. With the large
uncertainty associated with M$_{K_{S}}$, this consistent with that
of 2.5\,kpc inferred from its membership of Trumpler 27.     

\begin{figure}
\psfig{figure=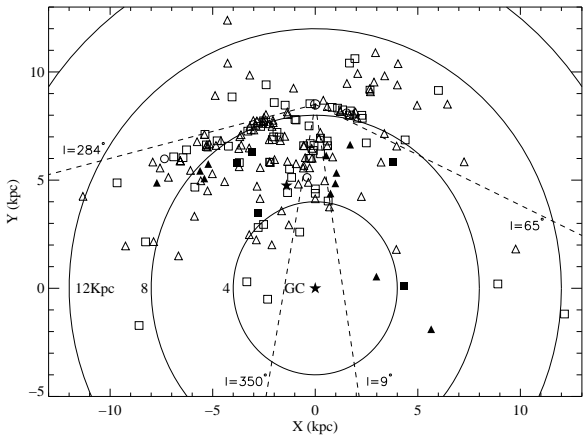,width=\columnwidth,angle=90.}
\caption{The Galactic WN (triangles), WC (squares) and WO (circles )
  distribution as projected onto the Galactic Plane.  Open symbols
  correspond to previously catalogued WR stars 
  \citep{derhucht01, derhucht06} whereas filled symbols represent the
  new WR stars presented here. The Sun ($\odot$) corresponds to a
  Galactrocentric distance of 8.5\,kpc and the symbol at
  the centre of the plot represents the Galactic Centre, Arches and
  Quintuplet clusters (N$_{TOT}$[WR]=60).  The 24 WR stars located
  within Westerlund~1 are also represented by a star and  have an
  assumed distance of 4.5\,kpc. The dashed lines indicates the
  latitude coverage of the GLIMPSE survey.}
\label{fig:distribution}
\end{figure}

In Figure~\ref{fig:distribution} we show the updated Galactic
distribution of WR stars, as projected onto the Galactic Plane.  We
confirm conclusions of \citet{conti90}, namely that the WR
stars appear to trace the spiral structure of the Galaxy, with one
spiral feature extending towards {\it l}$\sim$280$^{\circ}$.  Additionally,
the Galactic WR distribution shows remarkable similarities to that
of radio selected H{\sc ii} regions \citep[][their Fig~3]{russeil03}.  

The majority of the new WR stars are located between 4\,kpc $\lesssim
R_{G} \lesssim R_{\odot}$, the region where most of the previously
known WR stars are observed to cluster.  The exceptions are HDM\, 10,
12 and 13 which appear to be isolated from the bulk of the Galactic WR
stars.  This distribution suggests that our survey has identified WR
stars situated in more obscured regions of the inner Galaxy, rather than
those which are more distant from the Sun.

Recalling from Section~\ref{IR:Observations}, arbitrary magnitudes limits
of K$_{S} \sim$10.5 and K$_S \sim$11.0\,mag were placed on IR
spectroscopic observations to allow the maximum number of targets to
be observed.  This observational limit obviously influences the
distribution of the newly discovered WR stars shown in
Figure~\ref{fig:distribution} and the maximum distance out to which we
can detect WR stars is, of course, subtype dependant. For example, an
early-type WR star (M$_{K_{S}}$\,$\sim$\,--4\,mag) at a distance of 15\,kpc
would have a minimum apparent K$_{S}$ magnitude of 12\,mag,
well below the limiting magnitude of any of our followup observations
and the GLIMPSE [8.0\micron] completeness limit of 9.5\,mag.

Adopting an average K-band extinction of 0.86\,mag
(Table~\ref{tab:dist}) and a limiting magnitude of K$_{S}=$10.5\,mag,
our observations are capable of identifying early and late-WN stars
within $\sim$6.5 and 13.5\,kpc of the Sun, respectively.  Similarly
for WC stars, observations of WC5--8 and WC9 subtypes are limited 
to within 7.5 and 12\,kpc of the Sun, respectively.

If we were to extend our observations to cover the fainter targets in
our sample, then it is possible that we could probe the WR population
at larger distances.  However, this may prove to be difficult as a
combination of larger photometric uncertainties and crowding issues
may significantly affect the reliability of our selection criteria,
severely reducing the success rate of the survey.

\begin{table}
\begin{large}
\caption{Galactic distribution of WR stars as a function of
  Galactocentric distance. log(O/H)+12 is quoted for R$_{G}$ = 6.0,
  8.5 and 10\,kpc, assuming log(O/H)$_{\odot}$+12=8.66
  \citep{asplund04} and a metallcity gradient of \citet{esteban05}. }
\begin{center}
\begin{tabular}{lrrr}
\hline
R$_{G}$& $<$7.0&\multicolumn{1}{l}{7.0--10.0}&$>$10\\
log(O/H)+12&$\sim$8.8&8.66&$\sim$8.5\\
\hline
N$_{WN}$&96&68&13\\
N$_{WC}$&81&41&6\\
N$_{WO}$&1&2&0\\
N$_{WC}$/N$_{WN}$&0.84&0.60&0.46\\
N$_{WNE}$/N$_{WNL}$&0.4&3&2\\
N$_{WCE}$/N$_{WCL}$&0.01&1&2\\
\hline
\end{tabular}
\end{center}
\end{large}
\label{tab:distribution}
\end{table} 

It is well established that the absolute number and subtype
distribution of a WR population is dependant on metallicity
\citep{massey03}.  In the Milky Way, a slight metallicity gradient
($\Delta \log\mbox{(O/H)} =-0.044\pm0.01\mbox{dex\,kpc}^{-1}$
[\citet{esteban05}]) is reflected by a preponderance of late-type WC
and early-type WN stars at smaller galactrocentric distances.

In Table~\ref{tab:distribution} we list the number and subtype
distribution of Galactic WR stars according to our near-IR derived
distances.  This shows that the number of WC stars relative to WN
stars decreases with Galactocentric radius, confirming the earlier
conclusions of \citet{conti90} and \citet{derhucht01}.  Using oxygen
abundance as a proxy for metallicity, the early-to-late subtype
distribution is seen to be extremely sensitive to metallicity,
especially for WC stars. At Solar metallicity we count an equal number
of WCE stars relative to WCL stars whereas at higher metallicity WCL
stars significantly outnumber their early-type counterparts.  At
larger Galactic radii (lower metallicities) early-type stars dominate
the WC population with a WCE/WCL ratio of $\sim$2.  However, with only
small number statistics, completeness is one important issue that must
be considered when studying the WR distribution in the outer regions
of the Milky Way.  Similar conclusions are drawn for WN stars with
late-type stars being more common at smaller galactic radii, however
the range of observed ratios is much less extreme than that for WC
stars.  These results confirm previous conclusions drawn about the
Galactic WR distribution \citep{conti90, derhucht01} but highlight
how sensitive WR populations are to the metal content of their
environment.

Over the last two decades WR populations have been well sampled for a
variety of metallicty environments \citep{massey03}.  N(WC)/N(WN) is
observed to range from $\sim$0.1 in the metal-poor SMC
($Z$$\approx$0.2$Z_{\odot}$), whilst N(WC)/N(WN)$\sim$1 for M\,83.
For Solar metallicities, observations of the inner regions of M\,33 suggest
N(WC)/N(WN)$\sim$0.7, very similar to that of N(WC)/N(WN)=0.6 observed
for 7.0\,kpc$<R_{G}<$10\,kpc.  For the metal-rich inner Milky Way,
{\it Z} is equivalent to that of M\,83.  In M\,83, N(WC)/N(WN)$\sim$1
comparable to 0.84 obtained for  R$_{G}$$<$7.0\,kpc. 

\section{Non-WR detections}
\label{non:WR}

In order to better refine our selection criteria, it is vital to gain
an understanding of the objects that mimic the IR colours of WR stars.
In this section we will briefly comment on the objects which satisfied
our selection criteria, but turned out not to be WR stars. 


From spectroscopic follow-up observations, $\sim$75\% of spectra
displayed emission features \citep[c.f., 15\% in][]{homeier03}.  In
addition to the 15 WR stars, we have identified one previously
unreported planetary nebula and $\sim$120 (65\% of the total sample)
emission-line objects.  Of the 50  spectra (15\% of the total sample)
which did not contain emission features, the majority displayed strong
H absorption features consistent with mid- to late-type (A--G) dwarf
stars.


For the majority ($\sim$80 stars) of our emission-line objects,
emission features consisted of leading H transition lines; H$\alpha$
and H$\beta$ in our optical sample and P$\beta$ and Br$\gamma$ in our
IR sample.  

\begin{figure}
\begin{tabular}{c}
\psfig{figure=fig7a.ps,bbllx=45pt,bblly=60pt,bburx=550pt,bbury=790pt,clip=,width=\columnwidth,angle=-90.}\\
\psfig{figure=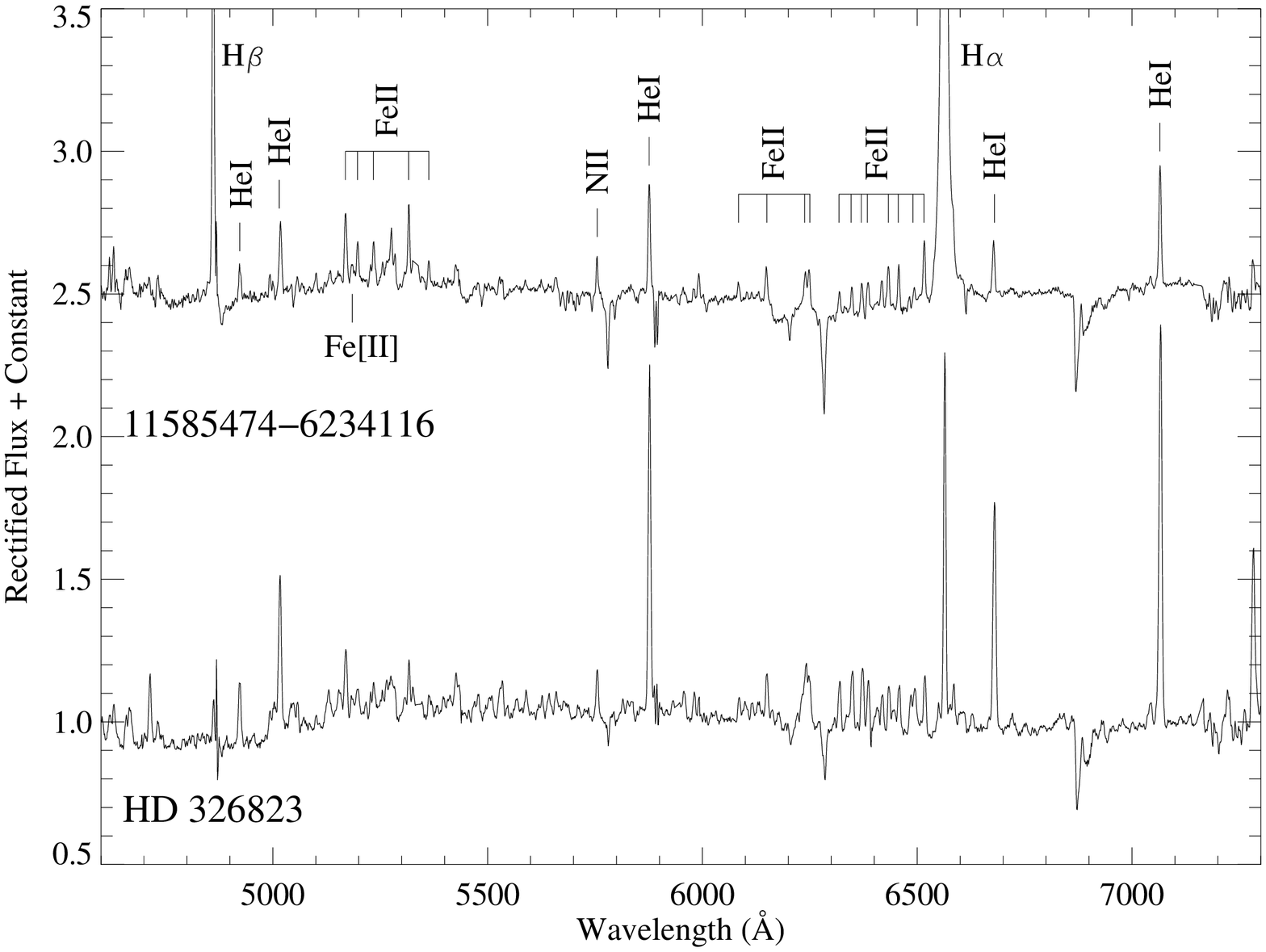,bbllx=75pt,bblly=65pt,bburx=540pt,bbury=715pt,clip=,width=\columnwidth,angle=90.}
\end{tabular}
\caption{Non-WR detections.  Top: A montage of CTIO/optical spectra of
  the H$\alpha$ emission-line objects detected.   Bottom: A spectral
  comparison of 2MASS source 11585474-6234116 with
  HD\,326823 (B[e]). The spectra have been been vertically offset (by
  1.5 continuum units) for clarity. Prominent emission features are
  marked. }
\label{fig:nonwr}
\end{figure}

In Figure~\ref{fig:nonwr} we show the rectified spectra
of a representative sample of these objects observed with the
CTIO/RCSpec.  
%
H$\alpha$ is always present in emission, whereas H$\beta$ is seen in
emission or absorption.  In addition to H$\alpha$ and H$\beta$, the
majority of objects display He\,{\sc i}\,$\lambda$6678
absorption/emission.  Helium and hydrogen features are characteristic
of early-type stars, but the absence of He\,{\sc ii}\,$\lambda$5411
is consistent with these objects being early B
supergiants/hypergiants.  Sources which do not show He\,{\sc i}
$\lambda$6678 must be later-type stars, possibly late B or early A
supergiants.

We have also identified a subset ($\sim$40) of objects which display
rich emission-line spectra, including H, He\,{\sc i} and numerous
lower-excitation metal lines, especially Fe\,{\sc ii}.  Objects which
display these emission-line features and near-IR excesses are usually
classified as peculiar B[e] stars.  A comparison of one these
objects with that of the B[e] star HD\,326823 is presented in
Fig~\ref{fig:nonwr}, clearly illustrating the similar spectral
morphologies.  Consequently, based solely from a spectral comparison
standpoint, we identify these objects as Be/B[e] candidates.

\section{Discussion \& Conclusions}
\label{conclusions}
Using a combination of near- and mid-IR colours we have identified
candidate WR stars in the Galactic Plane using the observed colours of
known Galactic WR stars as a ``training set''.  Initial spectroscopic
follow-up of 184 of our candidate WR stars has led to the discovery of
eleven new WN and four new WC stars.  

Up to now, WR searches have used narrowband K-band interference filters to
search for emission line excesses, making them particularly insensitive to
dusty WC stars.  Here, we have detected both WN and WC stars, with
none of the WC stars showing evidence for circumstellar dust.

The fifteen WR stars fall within 10$^{\circ}.2 < l <
331^{\circ}.0$. For fourteen of these, we estimate that $\mbox{R}_{G}
< \mbox{R}_{\odot}$ suggesting that this survey has been sampling WR
stars located in nearby, highly obscured regions of the Galaxy.  
Since an all sky survey was not possible and follow-up spectroscopy
was only capable of detecting the faintest WR subtypes to within
$\sim$6.5\,kpc, this survey is helping to complete the WR census
within the Solar vicinity.

Many recent WR surveys have concentrated on finding WR stars in
massive stellar clusters and with 30\% of the Galactic WR stars being
contained with four clusters \citep{derhucht06}, such searches are
very worthwhile.  Interestingly however, eight of the nine WR stars
detected in recent photometric surveys \citep{hopewell05, homeier03}
and the fifteen WR stars presented here are located in relatively
isolated regions.  Continually finding WR stars in isolation raises
many questions as to how they arrive there, are they runaway stars
expelled from clusters by supernovae events or binary/close
interactions, have their native clusters simply dispersed or do some
massive stars actually form in isolation?  Of course, with only small
number statistics we cannot attempt to address such issues here, but it
does highlight the importance of sample completeness and that searches
for WR stars in the Galactic Plane are extremely worthwhile.

In addition to WR stars, we have identified one new planetary nebula and
$\sim$120 H emitting sources within the Galaxy, most of which we
believe to be massive OB supergiants/hypergiants. 
With a WR yield rate of $\sim$7\%, and an massive star detection rate
of $\sim$65\%, our initial findings have proved that this survey is
one of the most successful searches for evolved massive stars in the
Galaxy.  However, since GLIMPSE is restricted to $>$$\pm10^{\circ}$
from the Galctic center, it is not possible that we will find
all of the missing WR stars, but will, in fact, complement other
ongoing WR searches.  

Our survey has revealed that using a combination of 2MASS and GLIMPSE
data it is possible to locate WR stars which had remained hidden from
view behind copious amounts of dust. Assuming that subsequent follow-up
observations of our remaining candidates achieves the same success
rate, we have the possibility of increasing the known Galactic WR
population twofold.

\section*{Acknowledgements} 
We wish to thank Michael Skrutskie and Michael Cushing for helping to
acquire and reduce some of the spectra presented here and Bobby Bus
for his invaluable assistance in showing us how operate SpeX.  We also
thank Paul Crowther for numerous, interesting discussions regarding
our spectra.

Some of the data presented here was acquired as a visiting astronomer
at the Infrared Telescope Facility, which is operated by the
University of Hawaii under Cooperative Agreement no. NCC 5-538 with
the National Aeronautics and Space Administration, Science Mission
Directorate, Planetary Astronomy Program.

This work is part of a {\it Spitzer} Cycle-2 archive programme funded
by JPL/Caltech and is part based on archival data taken by the {\it
Spitzer} Space Telescope.  This research has made use of the NASA/
IPAC Infrared Science Archive (IRSA).  {\it Spitzer} and IRSA are
operated by the Jet Propulsion Laboratory, California Institute of
Technology under contract with NASA.
\bibliographystyle{mn2e} \bibliography{abbrev,refs}

\appendix

\section{New WR spectra}

\begin{figure*}
\begin{tabular}{c}
\psfig{figure=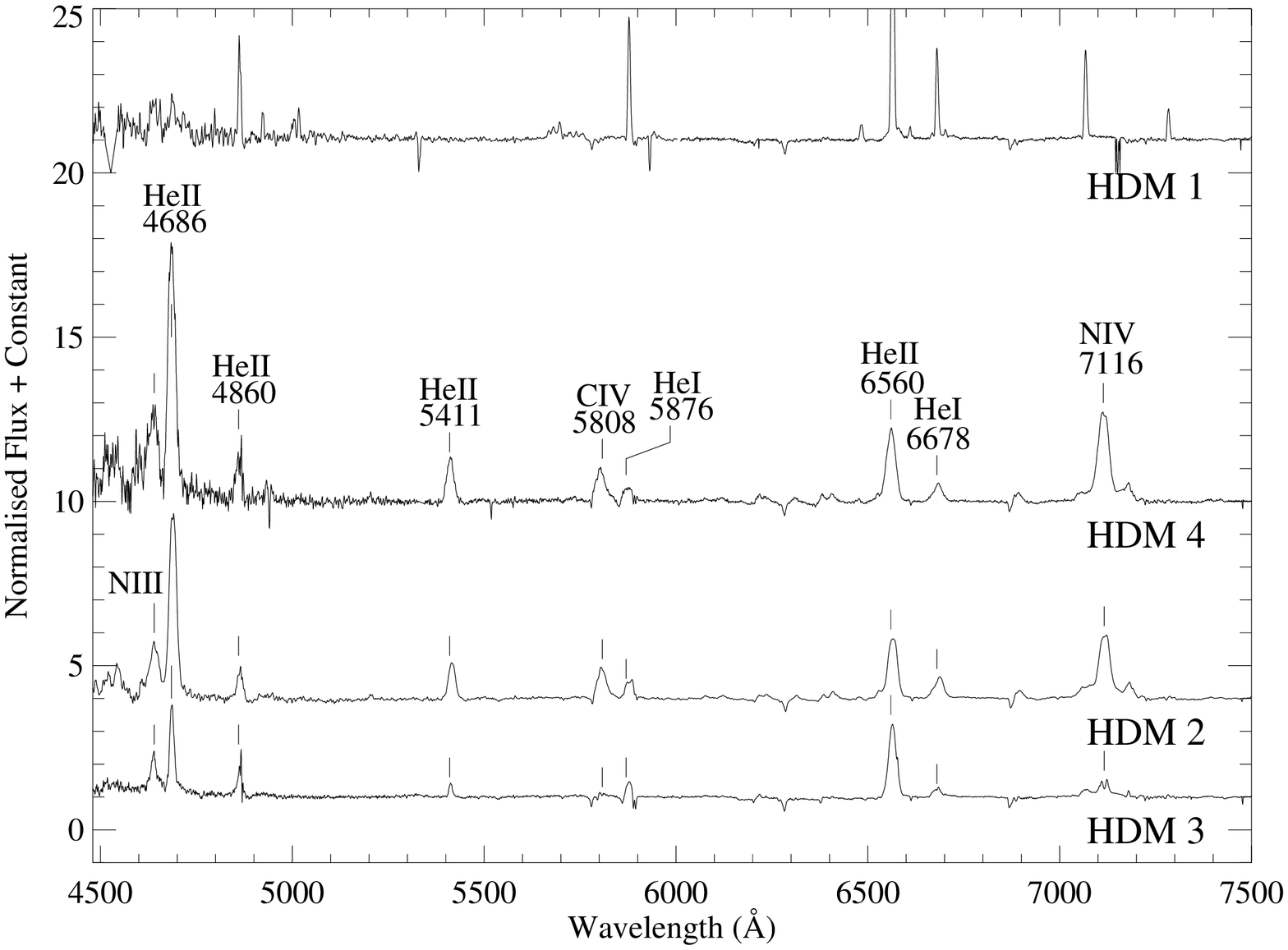,bbllx=60pt,bblly=60pt,bburx=540pt,bbury=740pt,clip=,width=15cm,angle=90.}\\
\psfig{figure=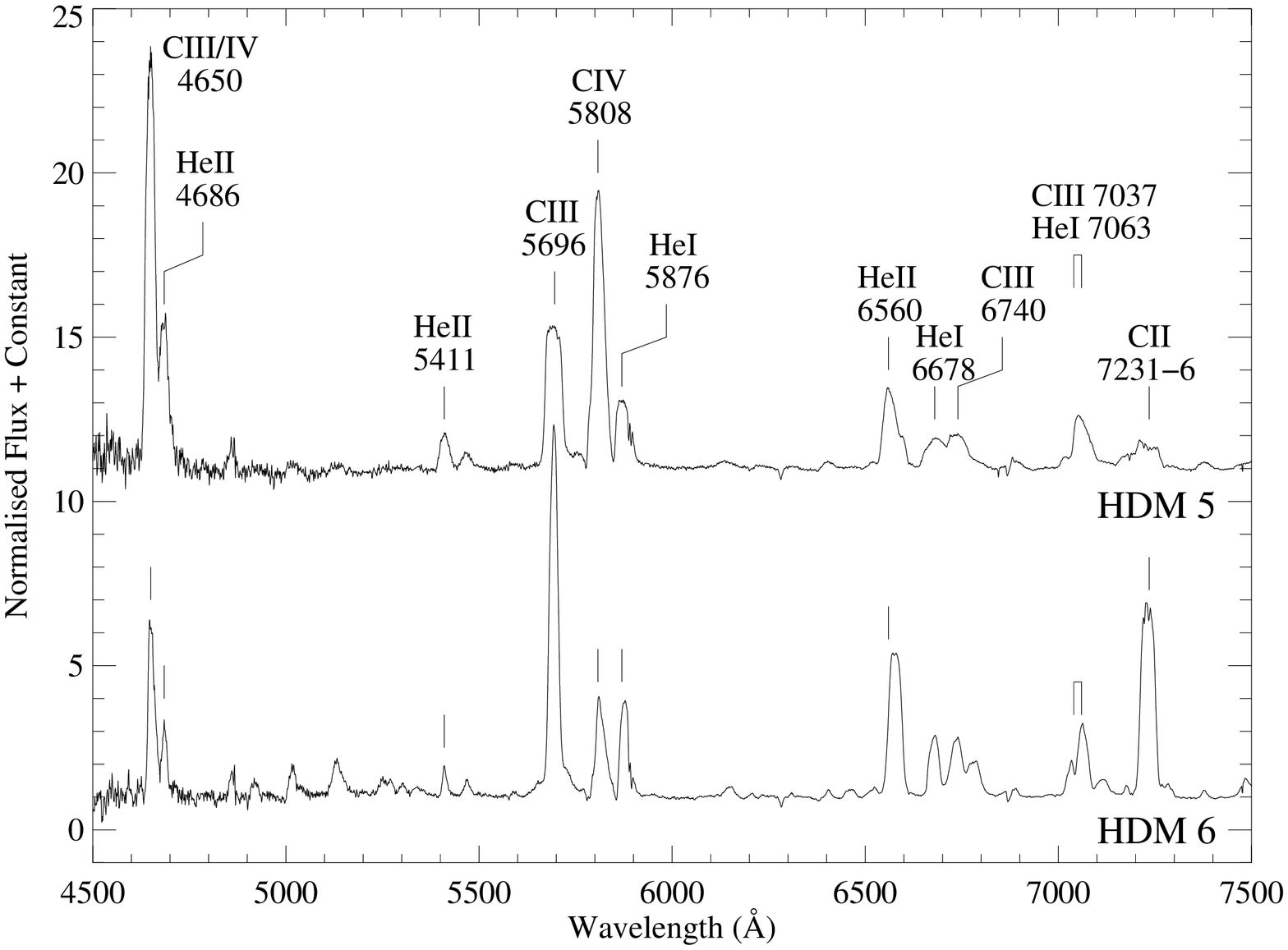,bbllx=60pt,bblly=60pt,bburx=540pt,bbury=740pt,clip=,width=15cm,angle=90.}\\
\end{tabular}
\caption{Optical CTIO/RCSpec data for six of the newly discovered WR
  stars.  WN stars are given in the top panel and WC star are
  presented in the bottom panel. }
\label{fig:spectra}
\end{figure*}

\begin{figure*}
\begin{tabular}{c}
\centerline{\psfig{figure=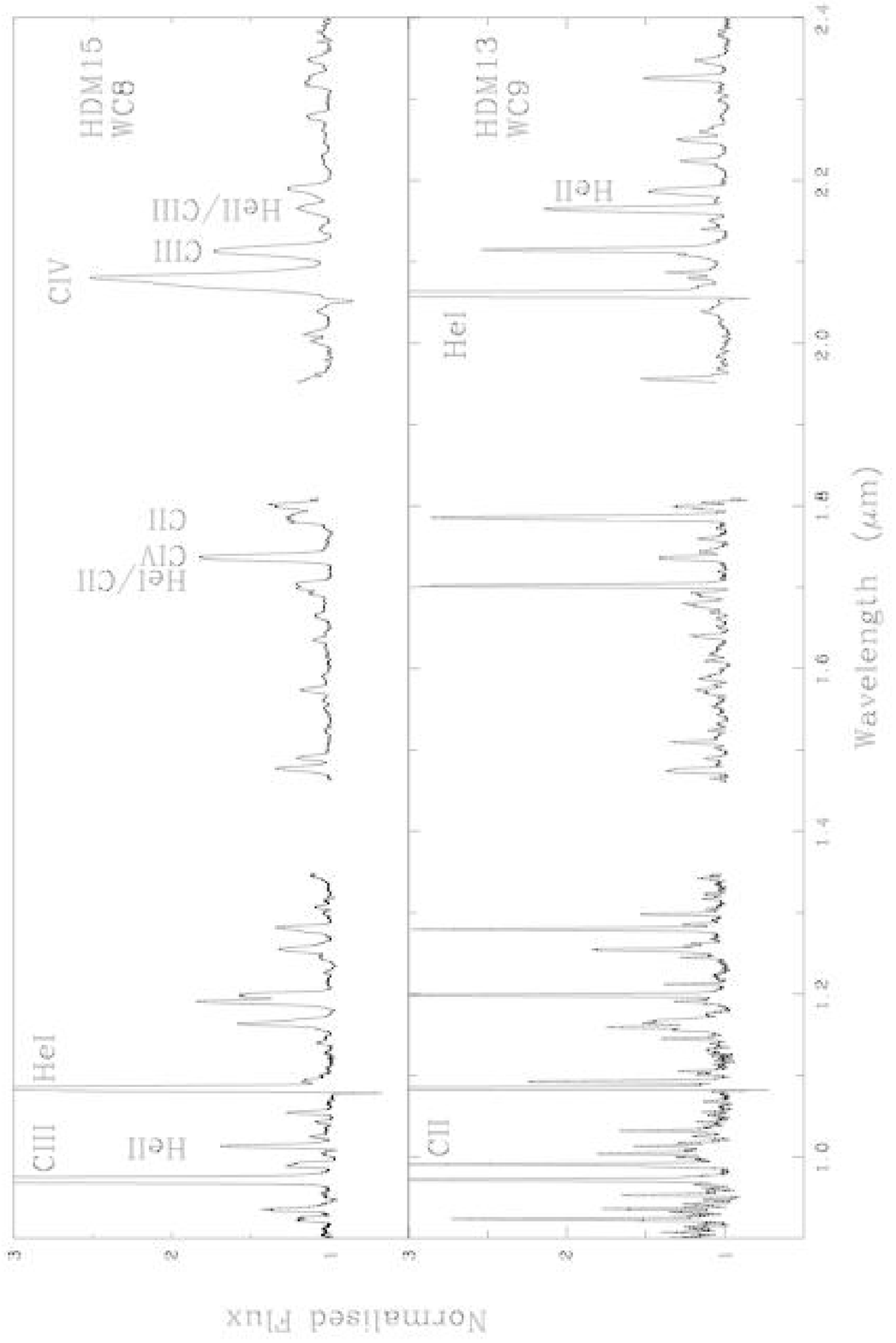,width=15cm,angle=-90.}}\\
\centerline{\psfig{figure=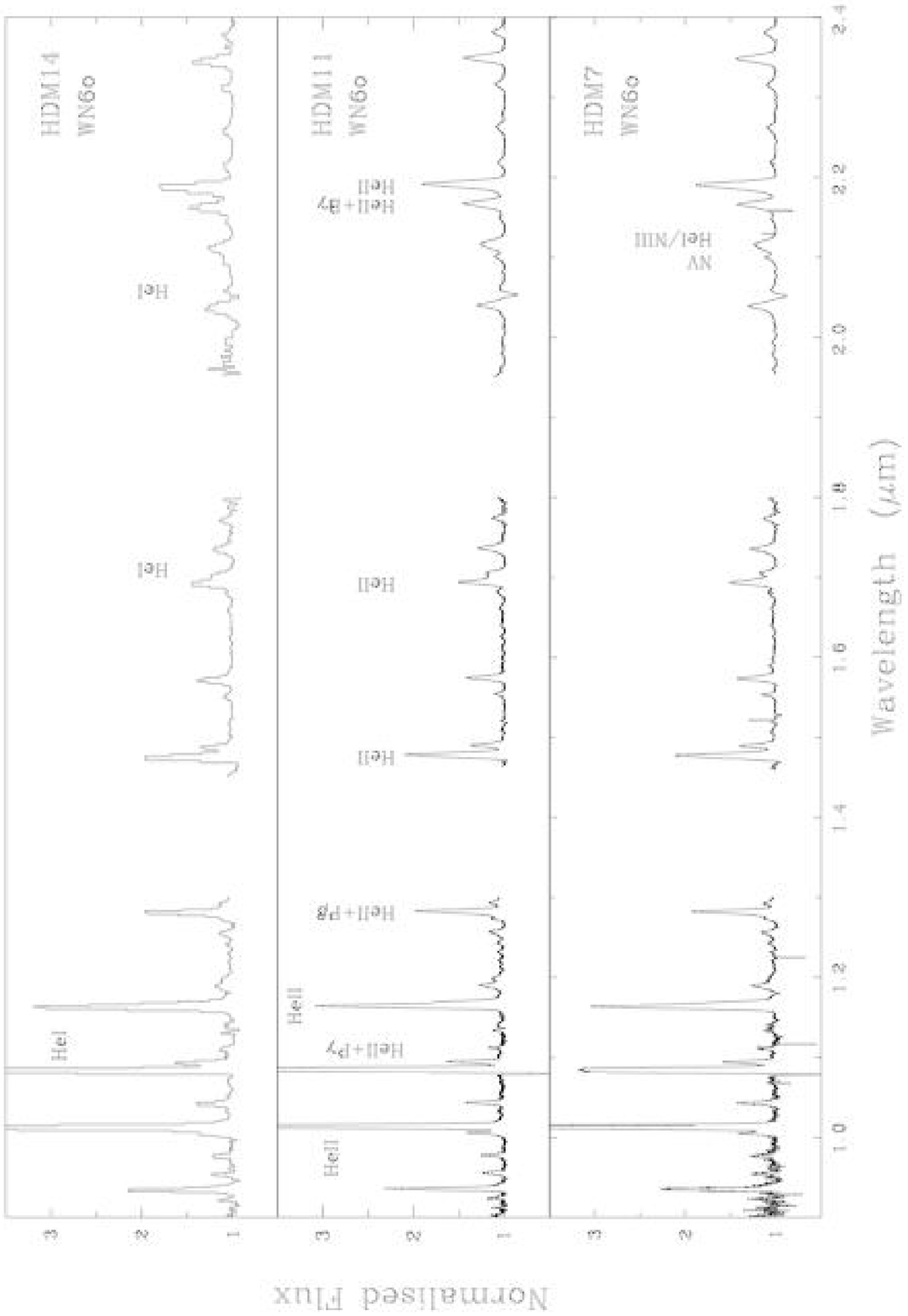,width=15cm,angle=-90.}}
\end{tabular}
\caption{Near-IR spectra of five of the newly confirmed Galactic WR
  stars.  WC stars are shown in the top panel and WN stars are bottom
  panel.  The spectra were acquired using IRTF/SpeX, except for
  HDM\,14 which was observed with APO/CorMASS.}\label{fig:spectra_ir}
\end{figure*}

\begin{figure*}
\begin{tabular}{c}
\centerline{\psfig{figure=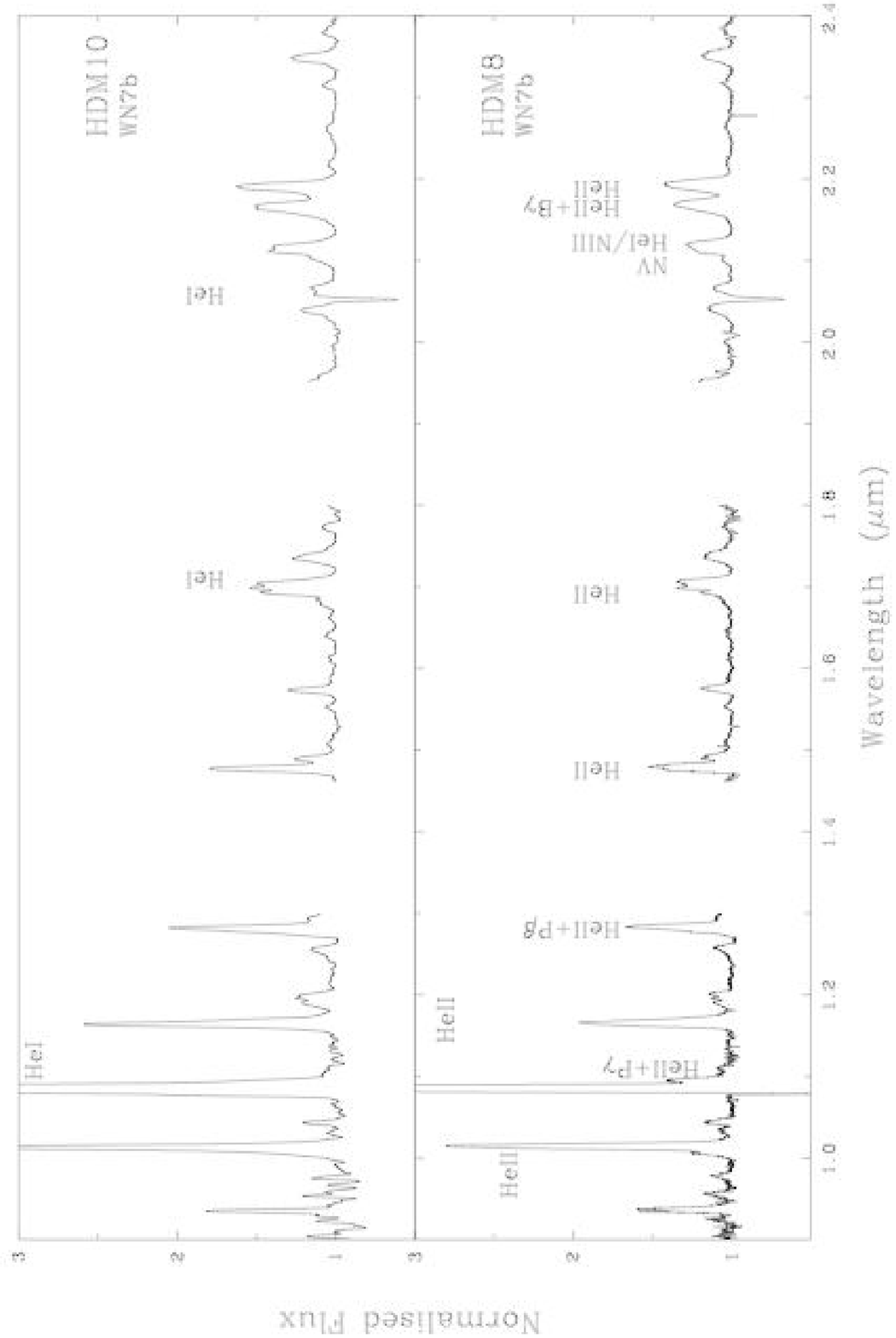,width=15cm,angle=-90.}}\\
\centerline{\psfig{figure=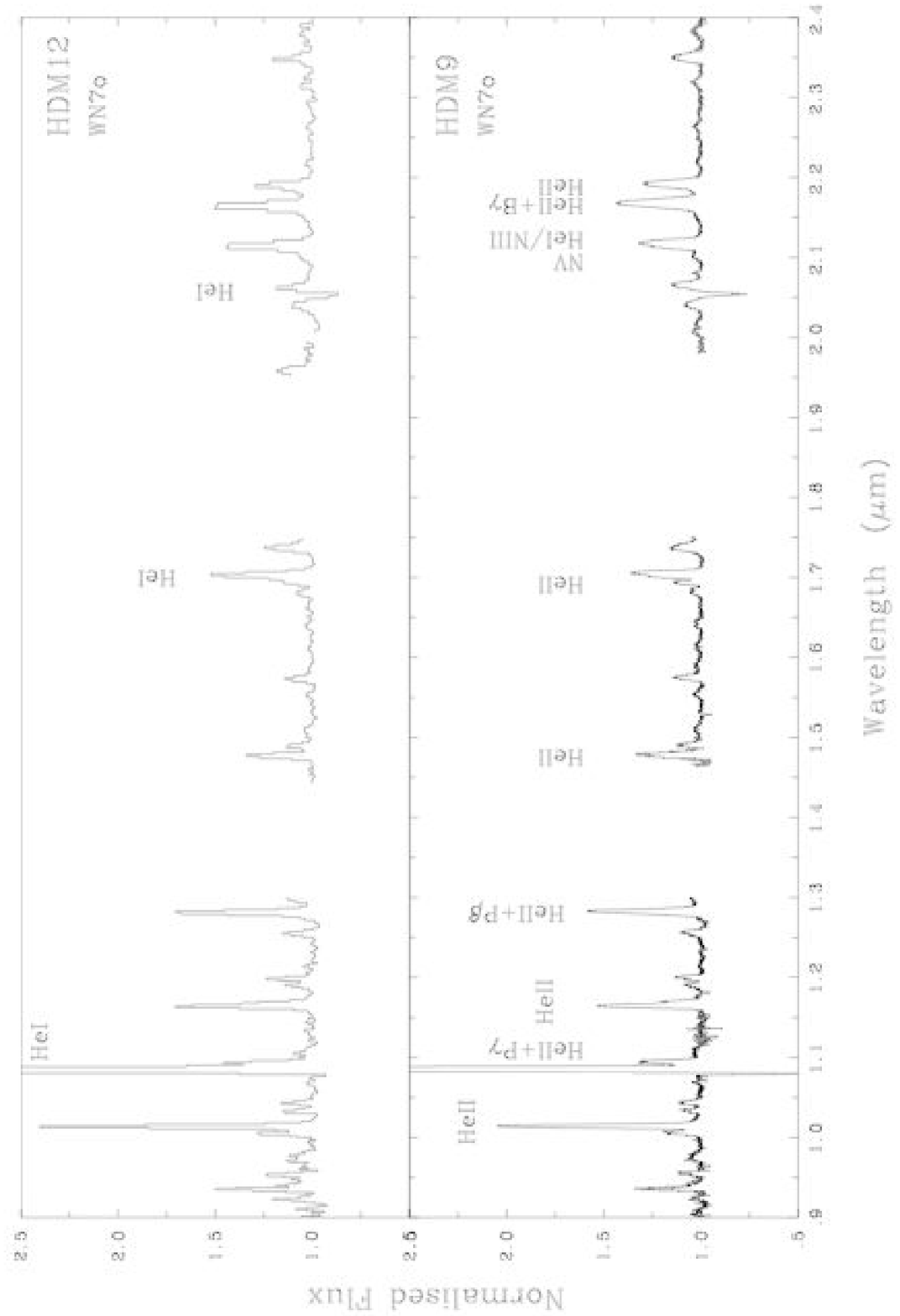,width=15cm,angle=-90.}}
\end{tabular}
\caption{Near-IR spectra of the four newly confirmed WN7 stars.
  Spectra have been grouped into broad (top panel) and narrow-lined
  (bottom panel) stars.  The spectra were acquired using IRTF/SpeX,
  except for HDM\,12 which was observed with APO/CorMASS.}
\label{fig:spectra_ir}
\end{figure*}

\section{Finding Charts}
\label{findingcharts}

\begin{figure*}
\centerline{\psfig{figure=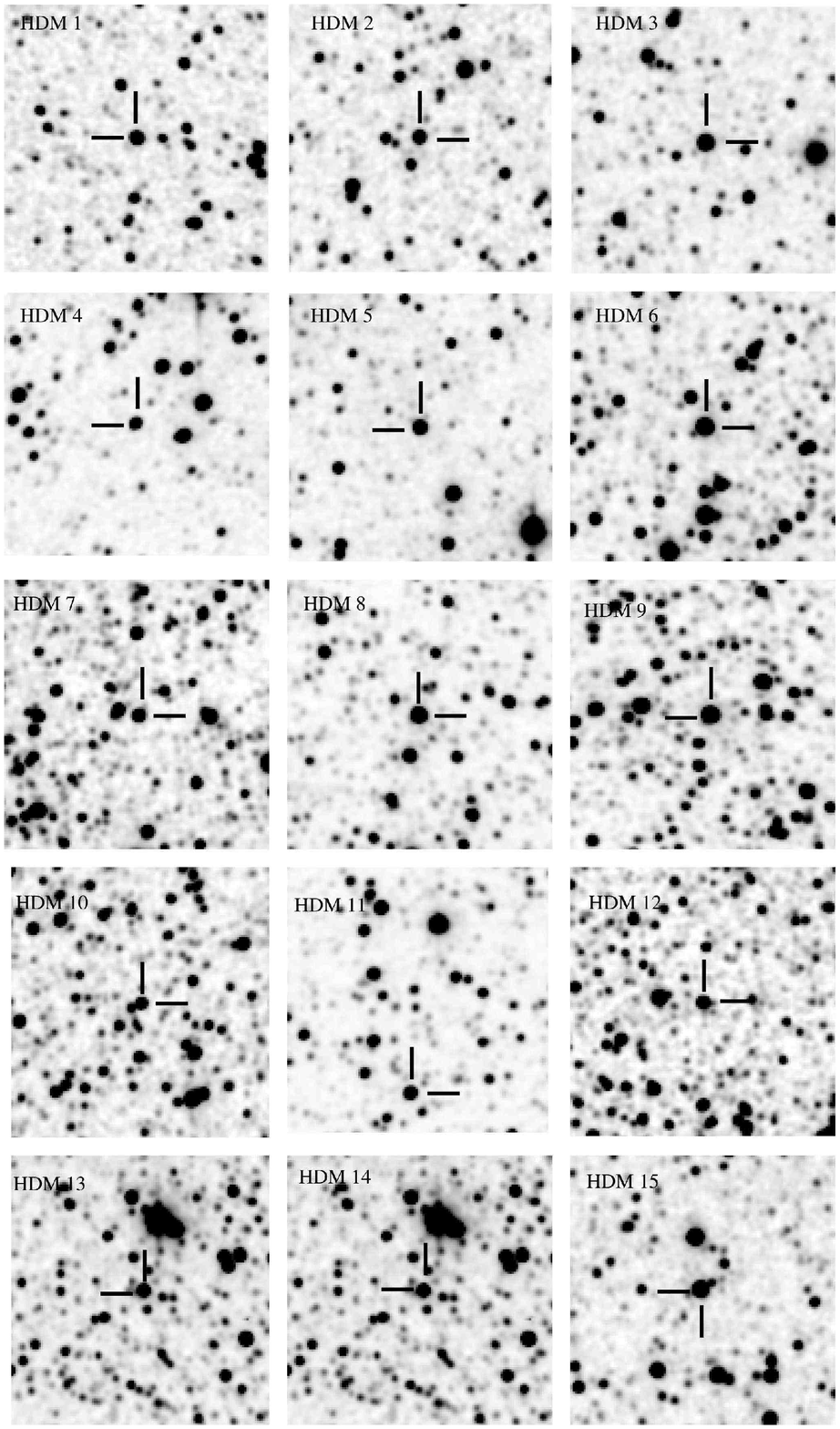,height=23cm,angle=0.}}
\caption{2 $\times$ 2\arcmin\ 2MASS K$_{S}$ finding charts for the
  newly discovered WR stars.  North is up and east is to the left on
  all images.   }
\label{fig:fc}
\end{figure*}

\end{document}